# Microscopy of hydrogen and hydrogen-vacancy defect structures on graphene devices

Dillon Wong[1,2,*], Yang Wang[1,2,*], Wuwei Jin[3], Hsin-Zon Tsai[1,2], Aaron Bostwick[4], Eli Rotenberg[4], Roland K. Kawakami[5,6], Alex Zettl[1,2,7], Arash A. Mostofi[3], Johannes Lischner[3], Michael F. Crommie[1,2,7,†]

[1] *Department of Physics, University of California at Berkeley, Berkeley CA, 94720, United States*

[2] *Materials Science Division, Lawrence Berkeley National Laboratory, Berkeley CA, 94720, United States*

[3] *Departments of Materials and Physics, and The Thomas Young Centre for Theory and Simulation of Materials, Imperial College London, London SW7 2AZ, United Kingdom*

[4] *Advanced Light Source, Lawrence Berkeley National Laboratory, Berkeley CA, 94720, United States*

[5] *Department of Physics and Astronomy, University of California, Riverside, California 92521, USA*

[6] *Department of Physics, The Ohio State University, Columbus, Ohio 43210, USA*

[7] *Kavli Energy NanoSciences Institute at the University of California, Berkeley and the Lawrence Berkeley National Laboratory, Berkeley, California 94720, USA*

\* These authors contributed equally to this work.
† Correspondence should be addressed to M.F.C. (Email: crommie@berkeley.edu)

## Abstract

We have used scanning tunneling microscopy (STM) to investigate two types of hydrogen defect structures on monolayer graphene supported by hexagonal boron nitride (h-BN) in a gated field-effect transistor configuration. The first H-defect type is created by bombarding graphene with 1-keV ionized hydrogen and is identified as two hydrogen atoms bonded to a graphene vacancy via comparison of experimental data to first-principles calculations. The second type of H defect is identified as dimerized hydrogen and is created by depositing atomic hydrogen having only thermal energy onto a graphene surface. Scanning tunneling spectroscopy (STS) measurements reveal that hydrogen dimers formed in this way open a new elastic channel in the tunneling conductance between an STM tip and graphene.

# I. INTRODUCTION

Hydrogen has been shown to be a useful element for modifying graphene's electronic properties. For example, hydrogen has been used to open a bandgap [1–3], enhance spin-orbit coupling [4], induce localization [5], and scatter electrons [6–8] in graphene. The interaction between hydrogen and graphene also makes graphene a good candidate for use in hydrogen storage [9] and related clean energy technologies. Moreover, the hydrogenation of graphene is believed to play a role in the formation of molecular $H_2$ and aromatic hydrocarbons in interstellar space [10]. Scanning tunneling microscopy (STM) measurements of hydrogen on graphene have revealed magnetism [11], insulating behavior [12], and reversible patterning [13,14]. Atomically resolved STM measurements of hydrogen adsorbed to gated graphene devices, however, has not yet been reported.

Here we present an STM study of hydrogen defect structures on gate-tunable graphene devices supported by hexagonal boron nitride (h-BN). Comparison of STM observations to density functional theory (DFT) calculations allows us to identify two species of hydrogen defects that occur when atomic hydrogen is deposited onto graphene/h-BN at room temperature: dihydrogen-monovacancies and hydrogen dimers. Dihydrogen-monovacancies appear when hydrogen ions are accelerated toward the graphene device through a 1 kV electric potential, while hydrogen dimers result from clustering of hydrogen atoms that are deposited onto graphene devices without any acceleration potential. Because hydrogen dimers change the local hybridization of the graphene lattice from $sp^2$ bonding to $sp^3$ bonding, a new elastic channel appears in the tunneling conductance between the STM tip and graphene. This reduces the relative contribution of phonon-assisted inelastic tunneling in scanning tunneling spectroscopy (STS) measurements of hydrogen dimers.

## II. METHODS

Our measurements were performed in an ultra-high vacuum (UHV base pressure = $10^{-10}$ torr) Omicron low-temperature (LT)-STM at T = 5 K using electrochemically etched platinum iridium (PtIr) and tungsten (W) tips calibrated against the Au(111) Shockley surface state. Differential conductance (d$I$/d$V$) was measured using the lock-in detection of the a.c. tunnel current modulated by a 6 mV rms, 613.7 Hz signal added to the voltage on the tip (-$V_s$). We used chemical vapor deposition (CVD) [15] to grow graphene samples which were then transferred onto h-BN flakes [16] on $SiO_2$/Si (see Ref. [17] for more details). h-BN provides a clean and inert substrate (compared to, e.g., graphite, metals, and silicon carbide) for investigating the intrinsic behavior of hydrogen defect structures on graphene [18–20]. The graphene samples were electrically contacted by Ti/Au electrodes deposited via stencil mask and the completed heterostructures were annealed at 400°C in UHV overnight. The charge carrier density in our graphene substrates was tuned during STM measurements by applying a back-gate voltage ($V_g$) on the heavily doped Si layer.

Molecular hydrogen was dissociated into atomic hydrogen for dosing onto graphene by passing $H_2$ gas through a tungsten tube held at an electric potential of 1 kV relative to a hot, thoriated-tungsten filament grounded to the UHV chamber (electron-beam bombardment of the tungsten tube heated it to T ≈ 1800 K). We verified that this procedure successfully produced atomic hydrogen by dissociating molecular deuterium ($D_2$) gas in front of a quadrupole mass analyzer. The atomic hydrogen was dosed for a few minutes onto room-temperature graphene with the chamber pressure rising to $10^{-7}$ torr. Hydrogen was dosed onto graphene device surfaces utilizing two different sample-biasing procedures. In procedure #1 the graphene sample was held at ground potential, thus allowing positive hydrogen ions to reach the surface with an average kinetic energy $\langle E_{KE} \rangle \approx 1000$ eV due to acceleration away from the positively biased tungsten tube. In procedure #2 the sample was biased with a deceleration voltage of 1 kV, the same bias as the tungsten tube. In this procedure positive ions reach the surface with an average kinetic energy commensurate with the temperature of the tungsten tube ($\langle E_{KE} \rangle \approx 0.2$ eV). In procedure #1 we annealed the graphene device at 400°C before transferring it to our T = 5 K LT-STM (this last anneal step was not done for procedure #2). As a control test we confirmed that none of the observations detailed below occur when hydrogen dosing gas is passed through a room-temperature tungsten tube instead of a heated tube.

## III. RESULTS AND DISCUSSION

The results presented in this paper are divided into two sections (A and B). In the first section we show our observations for hydrogen dose procedure #1 (where no deceleration voltage is applied to the sample). We show our experimental data for this dose procedure in Fig. 1 and the results of our theoretical modeling in Fig. 2. The second section of the paper shows our observations for hydrogen dose procedure #2 (where a deceleration voltage is applied to the sample). Figs. 3-4 show the STM data for this dose procedure, which are quite different from the experimental results of dose procedure #1. Fig. 5 shows the results of our theoretical modeling of the physical system that results from dose procedure #2. Overall, our results support the formation of dihydrogen-monovacancies on graphene/h-BN as a result of dose procedure #1 and adsorbed hydrogen dimers as a result of dose procedure #2.

### A. Dose procedure #1: Dihydrogen-monovacancies

Fig. 1 shows STM images of the graphene/h-BN surface after performing hydrogen dose procedure #1 (no deceleration voltage). Triangular-shaped defects can be seen in the larger area scan (Fig. 1a) with additional features visible in the atomically resolved close-up image (Fig.

1b).  As seen in Fig. 1b, each triangular defect is surrounded by a local $(\sqrt{3} \times \sqrt{3})R30°$ electronic superstructure and has a "lima-bean-shaped" (LB) object at its center.  Scanning over triangular defects with the STM tip causes the LB objects to occasionally rotate by 120° (this is demonstrated in Figs. 1c,d), but the triangular envelope surrounding each LB object does not rotate.  Gate-dependent d$I$/d$V$ spectroscopy on a triangular defect can be found in the Supplemental Material [21].

The experimental observations resulting from hydrogen dose procedure #1 are best understood by assuming that the triangular defects are monovacancies bonded to two hydrogen atoms.  Vacancies in graphene are known to create triangular modulations in the local density of states (LDOS) [22,23] and to cause intervalley scattering processes that lead to local $(\sqrt{3} \times \sqrt{3})R30°$ patterns [24], as seen in Fig. 1b.  This also explains why the LB objects can rotate under the influence of the STM tip: the dihydrogen-monovacancy structure has three degenerate configurations due to the $C_3$ symmetry of the vacancy.  The LB object rotates when a hydrogen atom jumps from one position on the vacancy to another, while the orientation of the surrounding triangle is fixed because it is $C_3$ symmetric and anchored to the sublattice of the missing carbon atom.  We note that we have only observed LB objects switching between two configurations (instead of three) (the mechanism for this is not clear but likely arises from asymmetry in the electric field of the tip [25] and the STM raster direction which cause a force with a well-defined direction to be exerted on LB objects).

The arguments in the previous paragraph do not rule out the possibility that only one H atom is bonded to the vacancy instead of a pair.  Indeed, previous *ab initio* simulations on monohydrogen-monovacancies do resemble our STM images [26–28].  In order to determine the structure of our triangular defects with more certainty we compared our experimental results to plane-wave pseudopotential density-functional theory (DFT) calculations carried out for monohydrogen-monovacancy as well as dihydrogen-monovacancy defect complexes using the QUANTUM ESPRESSO software package [29,30]. We employed a 6x6 graphene supercell, a 6x6 k-point grid, the Perdew-Burke-Ernzerhof (PBE) [29] generalized gradient approximation for exchange and correlation, ultrasoft pseudopotentials from the Garrity-Bennett-Rabe-Vanderbilt (GBRV) library (v1.2 for carbon and v1.4 for hydrogen) [31], and plane-wave energy cutoffs of 50 Ry and 500 Ry for the Kohn-Sham wavefunctions and the electron density, respectively. The periodically repeated graphene sheets were separated by 15 Å in the out-of-plane direction. We used Marzari-Vanderbilt smearing [32] with a smearing width of 0.005 Ry. Atomic positions were relaxed until the components of the forces on all atoms were less than $10^{-4}$ Ry/Bohr radius. We calculated STM images assuming $V_s$ = -1.36 V and a tip height of 3 Å.

Figs. 2a and 2b show simulated STM images resulting from our calculations for a monohydrogen-monovacancy and a dihydrogen-monovacancy, respectively. When one hydrogen is attached to the vacancy (Fig. 2a) the defect structure has an hourglass shape that is symmetric in the mirror plane parallel to the carbon-hydrogen bond. On the other hand, when two H atoms are present the simulated image (Fig. 2b) reproduces the lima-bean shape that we see in our topographic data. This suggests that the triangular defects are dihydrogen-monovacancies.

**B. Dose procedure #2: Hydrogen dimers**

*1. Experimental observations*

We now turn to the results of hydrogen dose procedure #2, where a deceleration voltage of 1 kV is applied to the graphene/h-BN sample during hydrogen dosing. Because this procedure generates no high-energy hydrogen ions (since the graphene sample and tungsten tube are held at the same bias potential), we expect the sample to experience less damage here than for hydrogen dose procedure #1. Fig. 3a shows a topographic image of graphene/h-BN after performing this modified hydrogen deposition process. Rounded rectangular protrusions can be seen randomly scattered throughout the surface. These features are completely removed by annealing the graphene device at 400°C. Some other structures were also seen, but the rectangular protrusions in Fig. 3a were the overwhelmingly dominant species (relative abundance > 95%) and no triangular defects were observed such as those seen after hydrogen dose procedure #1. The rectangular protrusions observed after dose procedure #2 were seen to align in three directions that correspond to the graphene substrate reciprocal lattice vector directions. This can be seen by comparing the blue dashed lines drawn through the long axes of the rectangular protrusions in Figs. 3a,c to the substrate crystallographic directions determined by taking the Fourier transform of an atomically resolved image of the bare graphene substrate (Fig. 3b). Fig. 3c shows a zoomed-in image of a typical rectangular protrusion which also reveals faint but discernible "legs" at the four corners.

In order to better understand the electronic properties of the rectangular protrusions arising from dose procedure #2, we probed them using gate-dependent $dI/dV$ spectroscopy. Figs. 4a-c show spectra acquired on a representative rectangular protrusion (red curves) compared to spectra obtained from the bare graphene substrate (black curves) for three different carrier concentrations: n-doped (Fig. 4a, $V_g$ = -20 V), neutral (Fig. 4b, $V_g$ = -30 V), and p-doped (Fig. 4c, $V_g$ = -40 V). The $dI/dV$ spectra on bare graphene (all taken for distances > 10 nm from a protrusion) each have a ~130 mV gap-like feature that arises from phonon-assisted inelastic tunneling [33,34]. The n-doped (p-doped) bare graphene spectrum has an additional local minimum below (above) the Fermi energy ($E_F$ at $V_s$ = 0 V) that is marked by a dashed green line.

These lines mark the location of the Dirac point for the doped graphene, whereas the Dirac point in the neutral case is at $V_s = 0$ V. The d$I$/d$V$ spectra measured on the rectangular protrusion do not show the gap feature at $V_s = 0$ V but do resemble the bare graphene d$I$/d$V$ curves in that they also exhibit a "V-like" shape (similar to what the bare graphene spectra would show if the ~130 mV gap were absent). However, the minimum of each "V" measured on the protrusion (highlighted by purple dashed lines) is shifted in energy relative to the local minimum measured on the bare doped graphene. For both n-doped and p-doped graphene the minima of the red curves are shifted $\hbar\omega \approx 60 \pm 10$ mV closer to the Fermi energy relative to the bare graphene Dirac points. For neutral graphene (Fig. 4b) the minimum of the red curve lies right at the Fermi energy and is coincident with the bare graphene Dirac point (which is somewhat obscured by the ~130 mV inelastic tunneling gap feature).

*2. Identification of hydrogen dimers*

These data allow us to identify the rectangular protrusions as hydrogen dimers that are believed to form via preferential sticking of hydrogen atoms near pre-existing chemisorbed hydrogen [35,36]. The elongation of the protrusions as well as the presence of the four "legs" (Fig. 3c) clearly break $C_3$ symmetry, so it is unlikely that these protrusions consist of single hydrogen atoms. It is far more likely that the rectangular protrusions are hydrogen dimers, which have previously been observed on graphite and graphene/SiC(0001) [35,37–39]. This hypothesis is supported by the fact that the protrusions are elongated along three equivalent directions parallel to the reciprocal lattice vectors obtained from the bare graphene Fourier transform (Fig. 3b inset). Since the nearest-neighbor bond directions and the primitive reciprocal lattice vectors are both rotated by 30° relative to the primitive real-space lattice vectors, it is reasonable to infer that the rectangular protrusions are each comprised of two hydrogen atoms sitting on either nearest-neighbor carbon atoms (the "ortho" configuration) or on carbon atoms located on opposite sides of a graphene hexagon (the "para" configuration). Figs. 5a,b show schematics depicting the relationship between the reciprocal-space vectors and the ortho and para geometries.

The ortho and para structures are believed to be the lowest-energy configurations for hydrogen dimerization on graphene, with ortho and para dimers having very similar total energies [38–40]. Hornekær *et al*. [38] have interpreted previous H/graphite STM data as indicating that ortho dimers manifest as elongated spheroids with a central node along their minor axis, while para dimers present as rectangular objects without a central node. Merino *et al*. [39], on the other hand, have interpreted previous H/graphene/SiC(0001) STM data as indicating that ortho dimers manifest as ellipsoids with no central node, while para dimers present as butterfly-shaped objects with a central node. Hornekær *et al*. and Merino *et al*. appear to disagree on whether it is the ortho or the para dimer that has or does not have a central node, while we do not observe any interior nodes in our rectangular protrusions (Fig. 3c).

To help distinguish between these different possibilities, we carried out DFT simulations using the same techniques as described above for our study of dihydrogen-monovacancies. Figs. 5c and 5d show simulated STM images (at $V_s = \pm 1.36$ V and tip height 3 Å – see Supplemental Material for simulations at other biases [21]) for an ortho dimer and a para dimer, respectively. A central node is present for both signs of the sample bias for the para dimer, while a central node is only seen for negative bias on the ortho dimer. This is in contrast to our experimental observations on the rectangular protrusions, in which we never see a central node for any sign of the sample bias. We do not know the origin of this discrepancy between theory and experiment, but it is possible that the shape of the tip apex smears the appearance of objects above the surface plane (tunneling between objects and the side of the tip apex also broadens the apparent lateral sizes of objects seen in topographic images), rendering us unable to resolve the central node (future scanned probe experiments involving chemically functionalized tips might be able to definitively identify the structure of these dimers [41–47]). We note, however, that the simulated para-dimer images (Fig. 5d) have four "legs" that strongly resemble our experimentally observed features (Fig. 3c), unlike the simulated ortho-dimer images (Fig. 5c). This suggests that our rectangular protrusions are para dimers, consistent with claims in Hornekær *et al*. and Merino *et al*. that the ortho dimers are ellipsoids and are not rectangular [38,39].

*3. Disappearance of the inelastic tunneling gap*

We now proceed to explain the d$I$/d$V$ spectra of Fig. 4. It is useful to first discuss in greater detail the origin of the ~130 mV gap-like feature at $E_F$ seen for bare graphene spectra (black curves in Fig. 4). These features arise since the STM tunneling conductance is dominated by an inelastic process in which electrons tunnel between the STM tip and graphene via virtual transitions to intermediate states near the Γ point in the graphene σ* band [33,34,48,49]. Since graphene's low-energy states are at K and K', this process must be accompanied by the emission of an out-of-plane K or K' phonon having $\hbar\omega_0 \approx 60$ meV in order to conserve crystal momentum. Inelastic tunneling only occurs above the threshold energy required to create a phonon, leading to an apparent gap of width $2\hbar\omega_0$ at the Fermi energy. This inelastic process is favored over direct tunneling into K and K' for our calibrated STM tips because the electronic states at Γ decay away from the graphene surface much more slowly than the states at K and K' [50].

The phonon-assisted inelastic tunneling gap is not present in the d$I$/d$V$ spectra obtained for hydrogen dimers (red curves in Fig. 4). This is most easily explained by the fact that chemisorbed hydrogen atoms change the local hybridization of carbon-carbon bonds from sp$^2$ to

sp$^3$ [4,51]. This breaks the translational symmetry of the crystal lattice, hence lifting the requirement for strict conservation of crystal momentum. Thus, phonon emission is no longer required to couple the Γ-point σ*-band states to the low-energy states at K and K', and the new defect-mediated elastic channel contributes more significantly to the tunneling current (a similar phenomenon has been observed previously for N impurities in graphene [52]). This interpretation is further supported by the observation that the minima of the red curves in Figs. 4a and 4c are shifted by an amount $\hbar\omega \approx \hbar\omega_0$ (60 meV) closer to the Fermi energy compared to the local minima of the black curves. The minima of the red curves represent the true energy location of $E_D$ because no phonon emission is required in the elastic channel, whereas the local minima of the black curves are located at energy $E_D \pm \hbar\omega_0$ since phonon emission is required for the inelastic channel. Hydrogen dimers thus suppress the phonon-assisted inelastic tunneling gap in graphene by opening a new elastic channel for electron tunneling. It is worth noting that the d$I$/d$V$ spectra of these hydrogen dimers do not show any peaks associated with carbon magnetism [11,22,53], which is consistent with theoretical expectations that ortho and para dimers are both nonmagnetic [40,54,55]. Theoretical calculations for the density of states (DOS) of ortho and para dimers can be found in Ref. [56] as well as in our Supplemental Material [21].

## IV. CONCLUSION

In summary we have used STM to image two types of hydrogen defect structures on graphene field-effect transistor devices that can be separately generated through the use of different hydrogen dosing parameters: hydrogen-vacancy complexes and H dimers. By comparing our experimental data to DFT simulations we have determined that the hydrogenated vacancies are likely composed of two H atoms, i.e. they are dihydrogen-monovacancies caused by bombarding the graphene surface with 1 keV ionized hydrogen (either protons or dihydrogen cations). Adsorbed H dimers, on the other hand, (which have no associated vacancy) suppress the phonon-assisted inelastic tunneling gap in graphene by opening a new elastic tunneling channel. These results provide new information on the types of hydrogen defect structures that can form on gated graphene devices at room temperature and should be useful for better understanding hydrogenated graphene and its potential applications.

This work was supported by the sp2-bonded materials program (KC2207) (STM measurements) funded by the Director, Office of Science, Office of Basic Energy Sciences, Materials Sciences and Engineering Division, of the US Department of Energy under Contract No. DE-AC02-05CH11231. The Molecular Foundry at LBNL was used for graphene characterization, which is funded by the Director, Office of Science, Office of Basic Energy Sciences, Scientific User Facilities Division, of the US Department of Energy under Contract No. DE-AC02-05CH11231. Support was also provided by National Science Foundation award DMR-1206512 (device fabrication, image analysis). W.J. acknowledges support from the EPSRC Centre for Doctoral Training in Theory and Simulation of Materials under Grant No. EP/L015579/1 (theory and simulation). J.L. and A.A.M. acknowledge support from the Thomas Young Centre under Grant No. TYC-101 (theory and simulation). This work used the Imperial College supercomputer cx1 and the ARCHER UK National Supercomputing Service through J.L.'s membership of the UK's HEC Materials Chemistry Consortium (funded by EPSRC (EP/L000202)). D.W. was supported by the Department of Defense through the National Defense Science & Engineering Graduate Fellowship Program, Grant No. 32 CFR 168a. We thank M.M. Ugeda, S. Coh, C. Chen, A.J. Bradley, and K.L. Meaker for helpful discussions and technical assistance. D.W. and Y.W. contributed equally to this work.

**Captions:**

FIG. 1. (a) STM topographic image of graphene/h-BN after bombarding the surface with 1 keV hydrogen ions (hydrogen dose procedure #1). Triangular defects with lima-bean-shaped (LB) centers are seen. (b) Zoomed-in topographic image of one triangular defect. (c) STM topographic image of a triangular defect before rotation of the LB center. (d) Same triangular defect as in (c) after rotating the LB center by 120° due to STM tip raster scan. Tunneling parameters: (a) $V_s$ = 500 mV, $I$ = 2 pA; (b) $V_s$ = 200 mV, $I$ = 500 pA; (c,d) $V_s$ = 250 mV, $I$ = 50 pA.

FIG. 2. (a) Ball-and-stick model adjacent to simulated STM image of a monohydrogen-monovacancy ($V_s$ = -1.36 V). (b) Same as (a) for a dihydrogen-monovacancy.

FIG. 3. (a) STM topographic image of graphene/h-BN surface after deposition of atomic hydrogen at room temperature (hydrogen dose procedure #2). Rounded rectangular protrusions can be observed on the surface in three different rotational configurations indicated by blue dashed lines. (b) Atomically resolved honeycomb lattice of graphene substrate in (a) and its Fourier transform (inset). (c) Zoomed-in topographic image of one rectangular protrusion with the color scale adjusted to reveal four "legs" positioned at the corners. The inset shows a line profile of the protrusion along the dashed blue line. Tunneling parameters: (a) $V_s$ = 500 mV, $I$ = 60 pA; (b) $V_s$ = -1 V, $I$ = 2 nA; (c) $V_s$ = 500 mV, $I$ = 60 pA.

FIG. 4. (a) d$I$/d$V$ spectrum of n-doped graphene substrate (black curve) and rounded rectangular protrusion (red curve). (b) Same as (a) for neutral graphene. (c) Same as (a) for p-doped

graphene. Initial tunneling parameters: $V_s$ = 500 mV, $I$ = 60 pA, 6 mV a.c. modulation; (a) $V_g$ = -20 V; (b) $V_g$ = -30 V; (c) $V_g$ = -40 V.

**FIG. 5.** (a) Sketch of graphene reciprocal space. The blue dashed arrows are primitive reciprocal lattice vectors and the orange hexagon is the graphene Brillouin zone. (b) Real-space representation of graphene. The black circles represent carbon atoms, the black lines are carbon-carbon bonds, and the green arrows are primitive lattice vectors. A pair of H atoms sitting on nearest-neighbor carbon atoms is called an ortho dimer, while a pair of H atoms sitting on opposite sides of a graphene hexagon is called a para dimer. The blue dashed lines connecting the H atoms in a pair are parallel to the reciprocal lattice vectors in (a). (c) Ball-and-stick model and simulated STM image of an ortho dimer (for $V_s$ = -1.36 V on the left and $V_s$ = +1.36 V on the right). (d) Same as (c) for a para dimer.

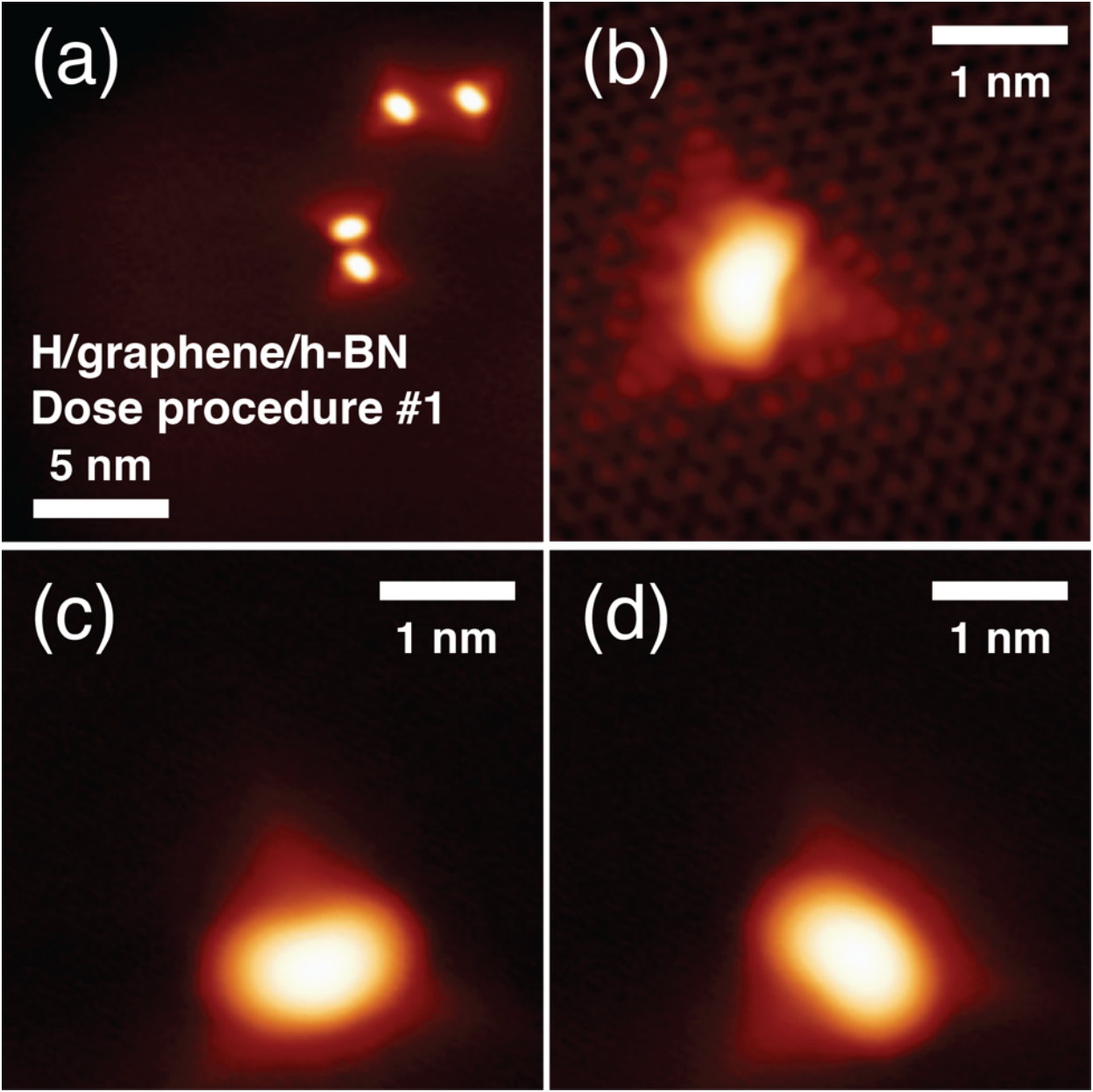

**FIGURE 1**

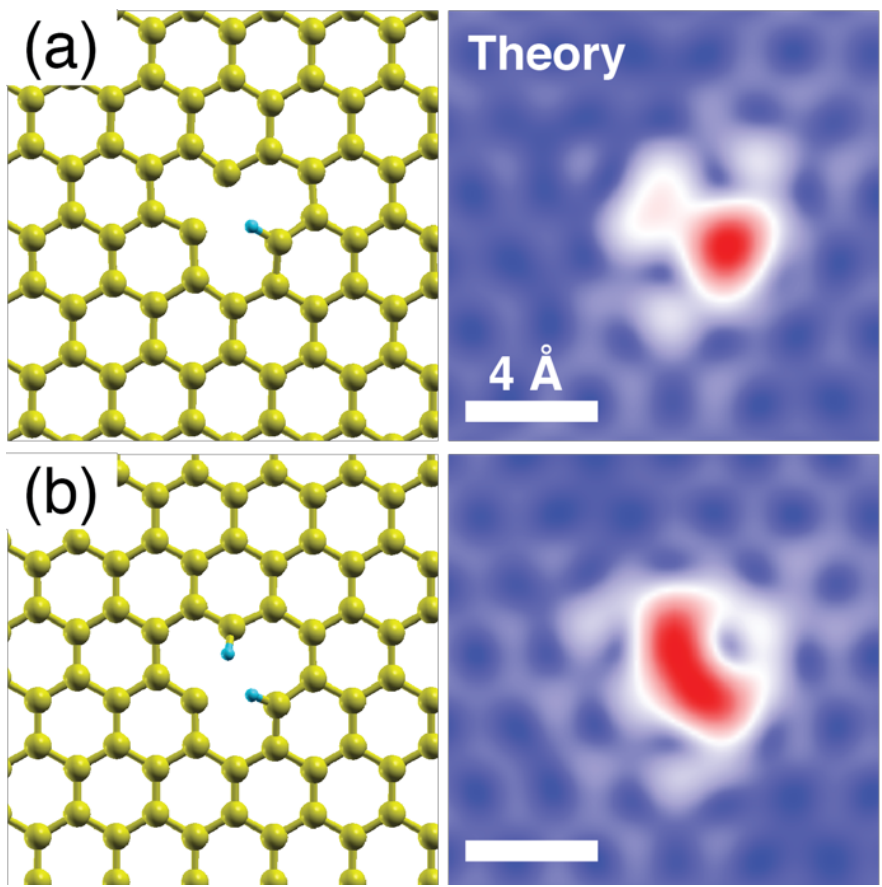

**FIGURE 2**

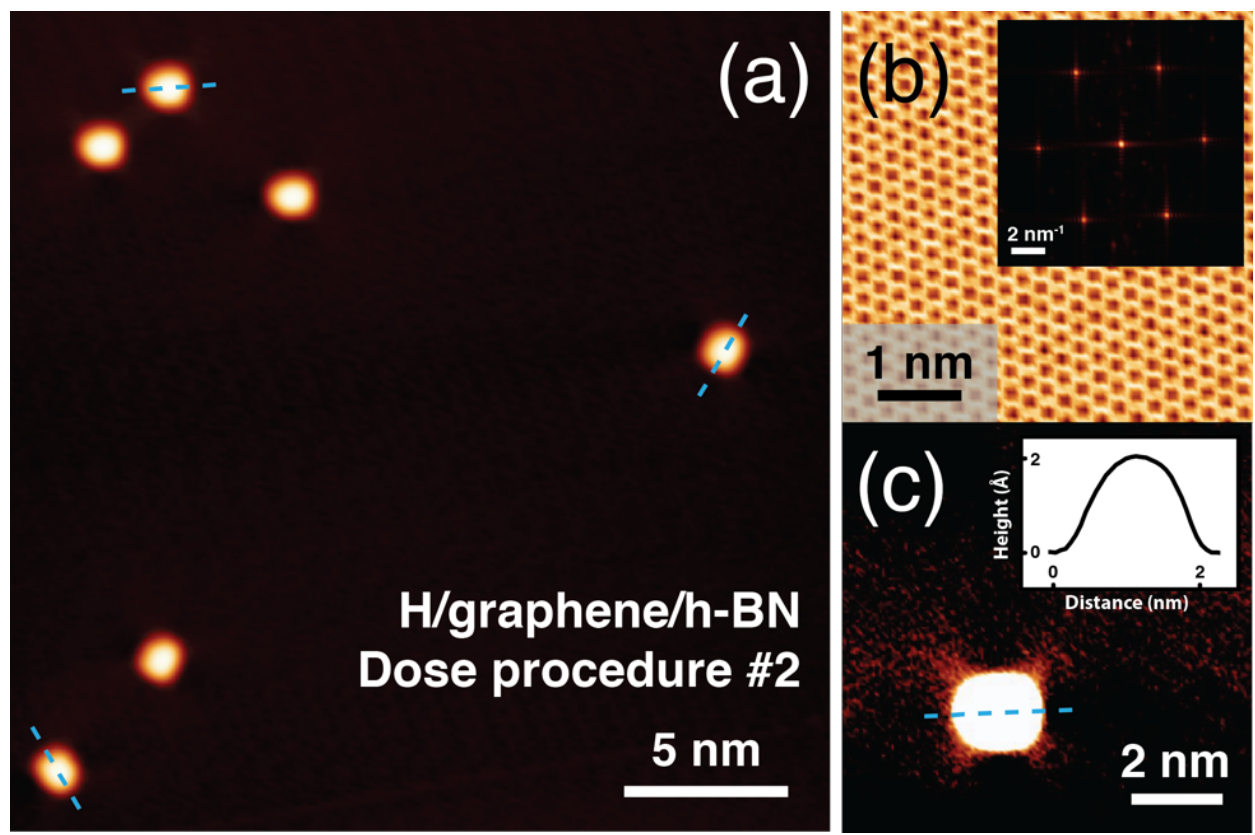

**FIGURE 3**

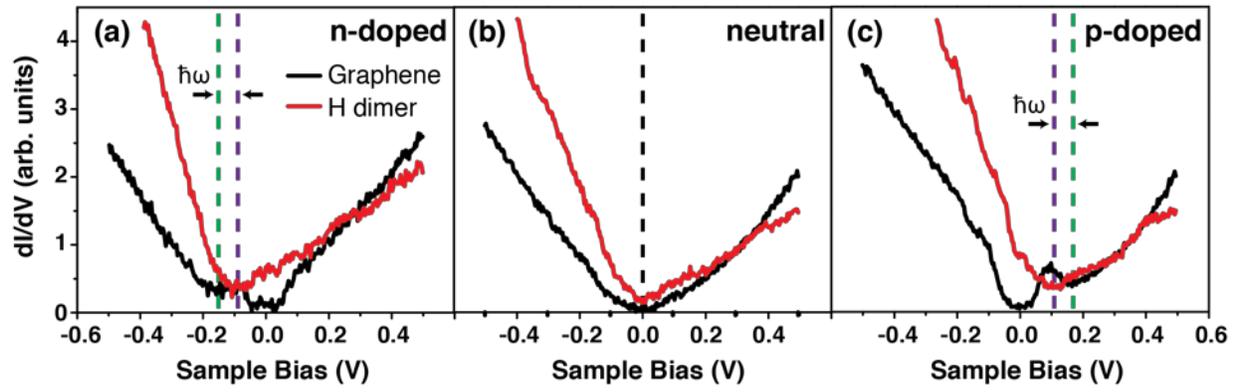

**FIGURE 4**

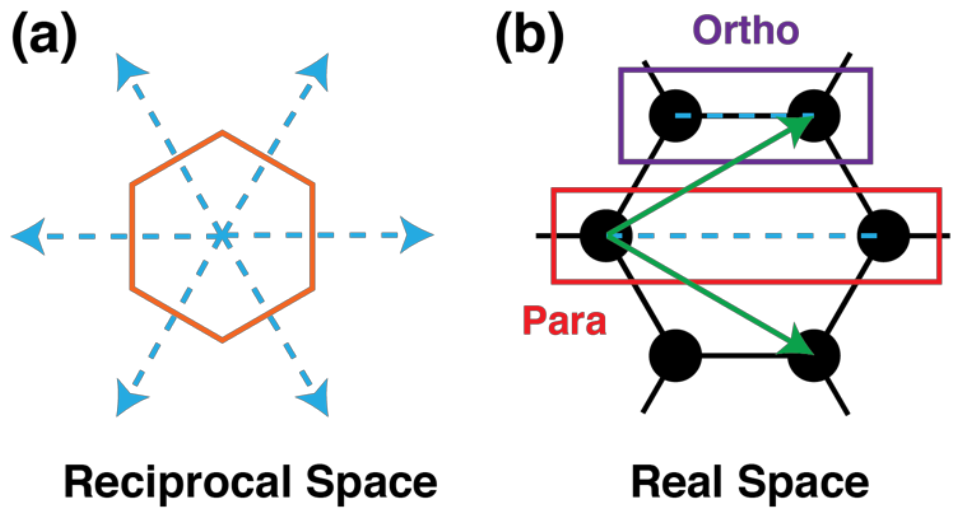
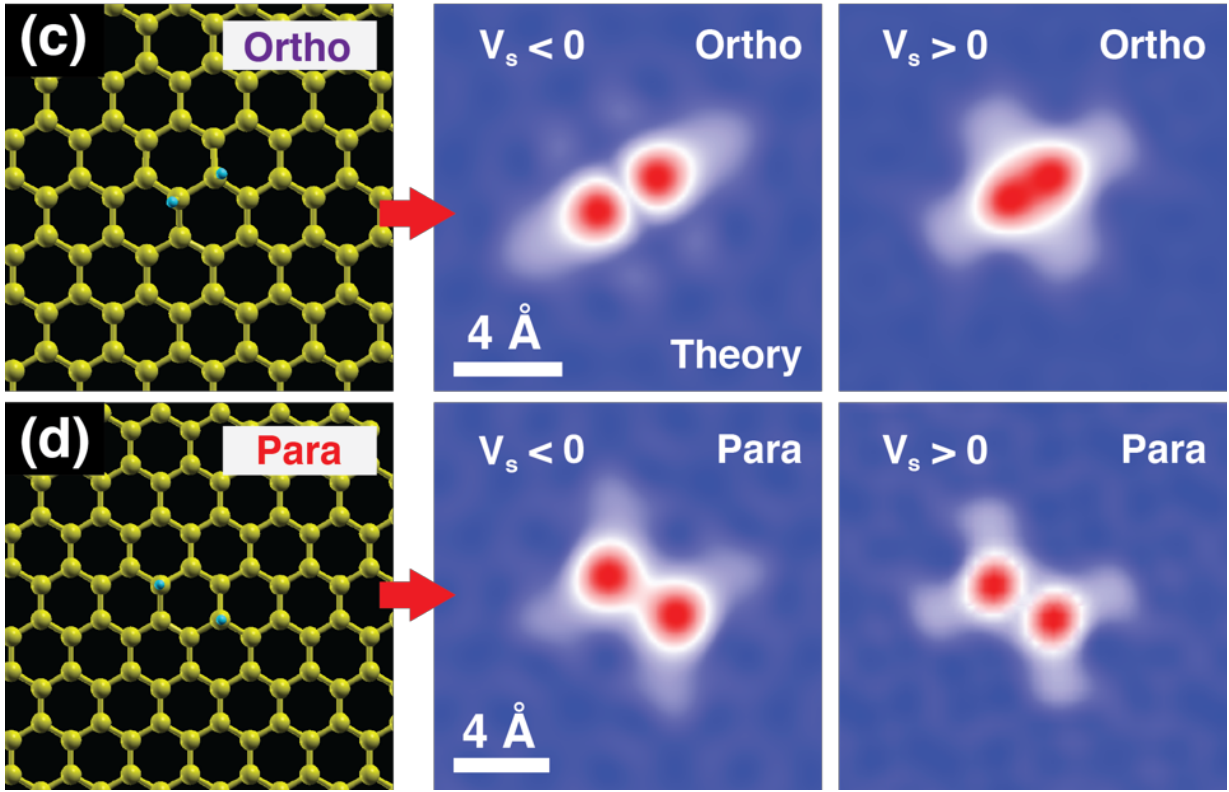

**FIGURE 5**